\documentstyle[prl,aps,twocolumn]{revtex}
\begin{document} 
\noindent
{\bf Neutron Scattering and $d$-Wave Symmetry in the Cuprates
}

\vspace{.1in}
In this comment we demonstrate that the general conjectures of Fong $et~ al$
on their ${\rm YBa_2Cu_3O_7}$ neutron experiments are realized in concrete
calculations of the
neutron scattering structure factor $S({\bf q} , \omega )$ in this material.
We make predictions
for similar experiments in ${\rm La_2Sr_{0.15}Cu_{1.85}O_4}$,
which yield a different q-structure due to the different fermiology.
Our calculations address both the
frequency and wave-vector dependence of the data and (1) include realistic
Fermi surface effects (2) are compatible with the normal state data
and (3) demonstrate the consistency of a $d$-wave order parameter,
in ${\rm YBa_2Cu_3O_7}$ although
they appear to be also compatible with an alternate symmetry
specific
to the orthorhombic bi-layer system.

We have shown \cite{us} that the normal state neutron data
 in both ${\rm YBa_2Cu_3O_{6+x}}$ and ${\rm  La_{2-x}Sr_xCuO_4}$ can be
understood as a competition between fermiology
and antiferromagnetic exchange effects $J({\bf q})$.  An infinite Coulomb U
expansion
leads to an RPA- like theory which yields an $S({\bf q} , \omega )$
 with wave-vector, ${\bf q} = (  q_x , q_y )$ [ as well as $q_z$ in the
bi-layer system] and $\omega$ dependences observed experimentally.
Similarly,
a four peak signature of the $d$-wave state at very low $\omega $ was
presented, although experiments on both cuprates have failed to confirm this
latter $d$-wave behavior.

Using our theory we plot the normal and superconducting data for
${\rm YBa_2Cu_3O_7}$ in Figure 1a as a function of
${\bf q} = (  q_x , q_y )$ at frequencies $\omega = 4
1meV$ slightly larger than twice the $d$-wave gap frequency
$2\Delta_o$.  The results are
in reasonable
agreement with Ref 1, although the calculated width is somewhat
wider than observed.
The dependence on ${\bf q} = (  q_x , q_y )$ is similar
for both $q_z = 0$ and
$q_z = \pi$, but the amplitude is only significant for the latter
wave vector\cite{bi}.  This follows in large part
because the two
$d$-wave gaps in each of the bilayer sub-bands are not appreciably
different\cite{dz}  on a scale of 41 meV. The mechanism for the formation
of the strong peak below ${\rm T_c}$ is , as conjectured in Ref 1 , a
consequence of
pair creation with one electron in  the + and the other electron in the -
lobe of the order parameter, whose wave vector
separation is near $(q_x,q_y)=( \pi , \pi )$. This enhances the weight around
$(q_x,q_y)=( \pi , \pi )$ in the Lindhard function. The further
inclusion of an antiferromagnetic exchange $J({\bf q})$ creates a strong peak
at $(q_x,q_y)=( \pi , \pi )$.
Comparable calculations are presented in
Figure 1b for  ${\rm La_2Sr_{0.15}Cu_{1.85}O_4}$, where the
Fermi surface nesting should lead to incommensurate d-wave peaks.
Here the characteristic
frequency is taken to be
17 meV slightly larger
than 16 meV for the estimated $2\Delta_o$.  While experiments have
been limited to lower $\omega$, future data should focus on this regime.

We end by noting that for the bi-layer system, the presence of
orthorhombicity can create a state\cite{dz} with a +$s$ gap on
one band and a -$s$ gap on the other, each of which gap is oriented along a
different
in-plane crystal axis.  This mimics a $d$-wave state and should
 also
be compatible with the neutron data of Ref. 1.  A similar state, however,
has no analogue in ${\rm  La_{2-x}Sr_xCuO_4}$.

This work
was supported by NSF-DMR-91-20000 and NSF-DMR-94-16926.

\vspace{.2in}
\noindent
Yuyao Zha$^a$  and  K. Levin$^b$\\

\noindent
$^a$Department of Physics,
University of Illinois,
1110 W. Green Street,
Urbana, IL 61801\\

\noindent
$^b$The James Franck Institute, The University of Chicago
5640 South
Ellis Avenue, Chicago, IL 60637
\begin{figure}
\caption{Calculated neutron scattering form factor Im$\chi$ as a function of
$(q_x,q_y)$ accouding to the model described in Ref\protect\cite{us},
for (a)${\rm YBa_2Cu_3O_7}$ by assuming the maximum of the superconducting
gap $\Delta_o
=20meV$, $J_0/J_c=0.7$; and (b)${\rm La_2Sr_{0.15}Cu_{1.85}O_4}$ for
$\Delta_o=8meV$, $J_0/J_c=0.6$.}
\end{figure}

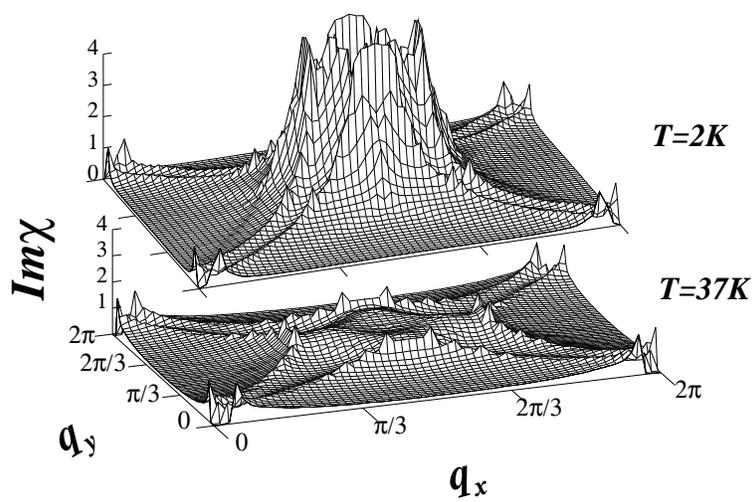
\end{document}